**Title:** Simultaneous achromatic and varifocal imaging with quartic metasurfaces in the visible

**Author List:**

Shane Colburn[1,*] & Arka Majumdar[1,2,*]

[1]Department of Electrical and Computer Engineering, University of Washington, Seattle, Washington 98195, USA.

[2]Department of Physics, University of Washington, Seattle, Washington 98195, USA.

*Correspondence to: scolbur2@uw.edu, arka@uw.edu

**Abstract:**

Two key metrics for imaging systems are their magnification and optical bandwidth. While high-quality imaging systems today achieve bandwidths spanning the whole visible spectrum and large changes in magnification via optical zoom, these often entail lens assemblies with bulky elements unfit for size-constrained applications. Metalenses present a methodology for miniaturization but their strong chromatic aberrations and the lack of a varifocal achromatic element limit their utility. While exemplary broadband achromatic metalenses are realizable via dispersion engineering, in practice, these designs are limited to small physical apertures as large area lenses would require phase compensating scatterers with aspect ratios infeasible for fabrication. Many applications, however, necessitate larger areas to collect more photons for better signal-to-noise ratio and furthermore must also operate with unpolarized light. In this paper, we simultaneously achieve achromatic operation at visible wavelengths and varifocal control using a polarization-insensitive, hybrid optical-digital system with area unconstrained by dispersion-engineered scatterers. We derive phase equations for a pair of conjugate metasurfaces that generate a focused accelerating beam for chromatic focal shift control and a wide tunable focal length range of 4.8 mm (a 667-diopter change). Utilizing this conjugate pair, we realize a near spectrally invariant point spread function across the visible regime. We then combine the metasurfaces with a post-capture deconvolution algorithm to image full-color patterns under incoherent white light, demonstrating an achromatic 5x zoom range. Simultaneously achromatic and varifocal metalenses could have applications in various fields including augmented reality, implantable microscopes, and machine vision sensors.

**Keywords:** Metasurfaces, Computational Imaging, Tunable Metasurfaces, Varifocal Lens, Achromatic Metasurface, Wavefront Coding

**TOC Graphic:**

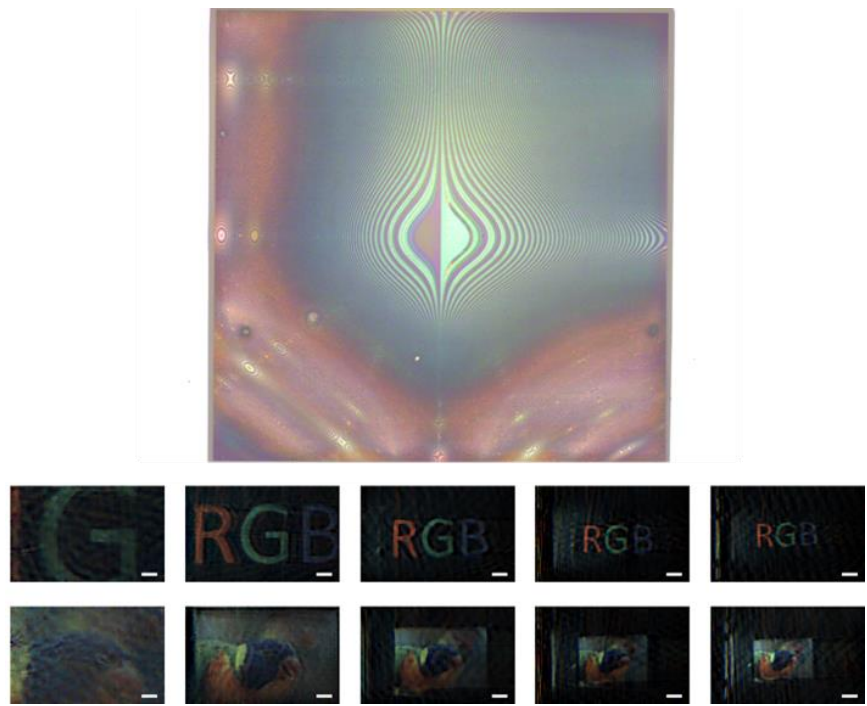

**Introduction:**

Imaging systems today boast high-quality achromatic imaging with optical zoom. These systems, however, often entail bulky elements incapable of meeting the demand for compact next-generation sensors and cameras. Computational imaging reduces this complexity by replacing sophisticated optics with simple elements and leveraging computation to transfer part of the imaging process into software[1–3]. This has enabled high-quality imaging using simple lenses combined with post-processing[4].

Separately, advancements in microfabrication and nanophotonics have driven development of metasurfaces, enabling miniaturization of optics by using quasi-periodic arrays of subwavelength scatterers to modify incident electromagnetic radiation[5–8]. By changing the amplitude, phase, and polarization of wavefronts, metasurfaces have enabled ultrathin lenses[9–18] (metalenses). A major limitation of metalenses, however, is their chromatic aberrations, which induce blur under broadband illumination. To mitigate this, recent works have added phase compensation or higher-order terms in the Taylor expansion of a metalens' phase function by utilizing dispersion-engineered scatterers[19–25]; however, while this is effective for achromatic focusing with certain small-aperture metalenses, it is not generalizable to large area elements[19,26]. As the required phase dispersion increases with radius even for a fixed numerical aperture[19], large area devices would require highly dispersive scatterers with exceedingly high quality factors or aspect ratios beyond what is achievable using state-of-the-art nanofabrication. Furthermore, until recently[24,25], many of these implementations relied on circular polarization to achieve achromatic focusing over a wide bandwidth[20,21,23], requiring additional polarizers and waveplates.

Recently, computational imaging and metasurfaces were used together to perform full-color imaging[27]. By introducing a cubic term to a metalens' phase profile, a spectrally invariant point spread function (PSF) was designed, enabling deconvolution with a single filter for achromatic imaging. At the cost of increased power consumption and latency from post-capture computation, this wavefront coding[28–30] technique circumvents the area scaling limitations of dispersion-engineered metalenses. These elements, like dispersion-engineered metalenses, however, are limited compared to state-of-the-art cameras in their inability to perform an optical zoom, which requires a varifocal lens. While various techniques exist for tuning metalenses, such as using MEMS to adjust the gap between two metalenses[31] or stretching metasurfaces on a flexible substrate[32–35], these approaches have not yet been demonstrated simultaneously with achromatic focusing. Though MEMS could feasibly support actuation of dispersion-engineered metalenses, it is limited to small devices to prevent dielectric breakdown, and while dispersion-engineered metalenses can be stretched, it is unknown whether the scatterers can maintain their dispersion as the distance between the scatterers changes. Recently, focal length tuning was also achieved with an achromatic response in the 483-620 nm range, but this required a continuous change of input polarization state[36].

To simultaneously achieve achromatic operation and an adjustable focal length, in this paper, we employ lateral actuation of a pair of wavefront-coded metasurfaces. Our design comprises two quartic metasurfaces that together produce a continuously tunable extended depth of focus (EDOF) lens, exhibiting a near spectrally invariant PSF at visible wavelengths for all focal lengths. In conjunction with a regularized post-capture deconvolution algorithm, we demonstrate white light imaging of colored patterns over a 5x zoom range. We report, to the best of our knowledge, the first polarization-independent metasurface system demonstrating simultaneous achromatic and varifocal zoom imaging.

**Results:**

*Design*

Our design is inspired by a traditional Alvarez lens[37–39], which comprises two cubic phase elements positioned in series along the optical axis. When the elements in an Alvarez lens are in alignment, the net phase delay is zero, as their spatial phase shifts exactly cancel each other according to

$$\theta_1(x,y) = -\theta_2(x,y) = A\left(\frac{1}{3}x^3 + xy^2\right), (1)$$

where $A$ is a constant and $(x, y)$ gives the in-plane position coordinates. On the other hand, when the two elements are laterally shifted in opposite directions, the superposition of their phase profiles yields a quadratic function. The focus of the quadratic function depends on the relative displacement, yielding a varifocal lens with a phase function

$$\varphi_{Alvarez}(x,y) = \theta_1(x+d,y) + \theta_2(x-d,y) = 2Ad(x^2+y^2) + \frac{2}{3}d^3, (2)$$

and a focal length

$$f(d) = \frac{\pi}{2\lambda A d}, (3)$$

where $d$ is the lateral displacement and $\lambda$ is the wavelength. If the phase elements are implemented as metasurfaces, however, the system would exhibit a significant chromatic focal shift, typical for diffractive lenses. Its point spread function (PSF) is strongly wavelength-dependent and yields zeros for large spatial frequency bands in the corresponding modulation transfer functions[28,29] (MTFs) for wavelengths deviating from the designed value. As was recently demonstrated, however, the addition of a cubic term to a metalens' phase profile provides an EDOF that yields a PSF that is nearly insensitive to longitudinal chromatic focal shift[27]. Instead of focusing to a point like a metalens, such an element produces a focused Airy beam and an asymmetric PSF that blurs images. Via a monochromatic PSF calibration measurement, however, this blur can be negated via deconvolution across the visible spectrum. We emphasize that such a deconvolution approach cannot be applied for the case of a singlet metalens under white light illumination, as the wavelength dependence of the PSF and the large increase in PSF size results in a loss of higher spatial frequency information that deconvolution cannot recover[27]. Modifying equation (2) to include a cubic term, neglecting the $d^3$ constant phase, and substituting the focal length from equation (3), the phase

$$\varphi_{EDOF} = \frac{\pi(x^2+y^2)}{\lambda f} + \frac{\alpha}{L^3}(x^3 + y^3), (4)$$

simultaneously imparts the profile for a focusing lens and extends its depth of focus. Here, $L$ denotes half the aperture width and $\alpha$ is the cubic phase strength (i.e., the number of $2\pi$ cycles from the cubic phase term when traversing a path from the origin to the aperture edge in the $x$ direction). Expressing $\varphi_{EDOF}$ as a superposition of two oppositely signed and laterally displaced phase functions,

$$\varphi_{EDOF} = \theta_{plate}(x + d, y) - \theta_{plate}(x - d, y), (5)$$

we can relate the derivative of $\theta_{plate}$ to this difference and $\varphi_{EDOF}$ via the two-sided definition of the derivative as below

$$\frac{\partial \theta_{plate}}{\partial x} = \lim_{d \to 0} \frac{\theta_{plate}(x+d, y) - \theta_{plate}(x-d, y)}{2d} = \lim_{d \to 0} \frac{\varphi_{EDOF}}{2d}, (6)$$

Making the substitution $\frac{\alpha}{L^3} = Bd$, where $B$ is a constant, we can find

$$\theta_{plate}(x, y) = A\left(\frac{1}{3}x^3 + xy^2\right) + B\left(\frac{1}{8}x^4 + \frac{1}{2}xy^3\right), (7)$$

When $B = 0$, $\theta_{plate} = \theta_1$ and the superposition of the two laterally displaced plates behaves exactly as an Alvarez lens; however, for nonzero B, $\theta_{plate}$ becomes a quartic phase polynomial in $x$. Continuous lateral displacement of these conjugate plates yields a tunable focal length EDOF lens with a phase distribution given by equation (4). Instead of focusing to a point, this system produces a tunable focused accelerating beam that generates spectrally invariant point spread functions that enable wavelength-uniform deconvolution[27] for each focal length. We refer to this device as an EDOF Alvarez metalens henceforth. While tunable cubic phase masks via conjugate quartic phase plates have been theoretically studied before[40–42], our design is the first to our knowledge that combines the functionality of both tunable focusing and tunable wavefront coding with only two elements.

To implement the phase profile of equation (7) and its inverse with metasurfaces, we choose silicon nitride as the material for our scatterers due to its broad transparency window and CMOS compatibility[17,43,44]. The scatterers are polarization-insensitive cylindrical nanoposts arranged in a square lattice on a quartz substrate (Figure 1A-B). The phase shift mechanism of these nanoposts arises from an ensemble of oscillating modes within the nanoposts that couple amongst themselves at the top and bottom interfaces of the post[12,45,46]. By adjusting the diameter of the nanoposts, the modal composition varies, modifying the transmission coefficient. At a nominal wavelength of 530 nm (green light), we simulate the transmission coefficient of a periodic array of these nanoposts with a thickness $t = 600\ nm$ and period $p = 350\ nm$ (Figure 1C) (see supporting information, Figure S1 and S2, for simulations at other wavelengths and justification of the local phase approximation respectively). In sweeping the diameter of the nanoposts, the transmission phase exhibits a nearly $2\pi$ range with amplitude close to unity except for several resonances that are avoided when selecting diameters for phase shifting elements.

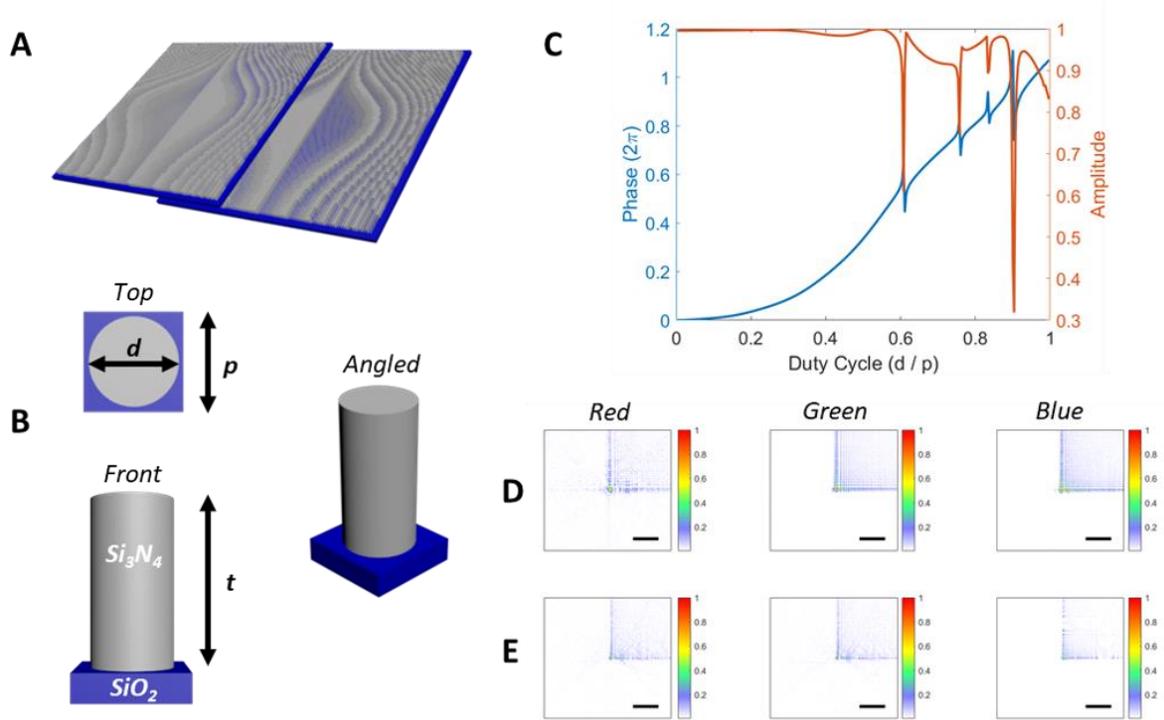

Figure 1: **Metasurface Design and Simulation** Schematics are shown of the Alvarez EDOF metalens system (A) and the nanoposts that comprise the metasurfaces (B). (C) The simulated amplitude and phase of the transmission coefficient of a periodic array of the designed nanoposts. Simulated point spread functions are shown for lateral displacements of 50 µm (D) and 100 µm (E) for the designed system at image planes of 3 mm and 1.5 mm respectively. Scale bar 80 µm.

To change the focal length from 6 mm to 1.2 mm by tuning $d$ from 25 µm to 125 µm at 530 nm wavelength, we use a value of $A = 1.98 \times 10^{13} \ m^{-3}$. A lower value of $A$ increases the focal length but reduces the numerical aperture for a fixed metasurface size. The value of $B$ depends on the desired operating wavelength range, which here spans from blue (455 nm) to red (625 nm) light. Higher values of $B$ increase the depth of focus at the cost of reducing the signal-to-noise-ratio (SNR). As the range of chromatic focal shift increases with a wider optical bandwidth, however, the focal depth must be extended to ensure a spectrally invariant PSF over the whole wavelength range. For our wavelength range, $B = 7.59 \times 10^{16} \, m^{-4}$ satisfies this requirement (Figure S10 shows the corresponding nanopost radii distributions for the resulting metasurfaces). With an aperture of 1 mm, Figure 1D compares the simulated PSFs of our system under incoherent red, green, and blue illumination. We use a fixed displacement of $d = 50 \, \mu m$ at an image plane of 3 mm, demonstrating highly similar responses for each wavelength. Figure 1E examines PSFs of the system at the same three wavelengths but with a displacement of $d = 100 \, \mu m$, producing near-invariant PSFs at a 1.5 mm image plane. The difference in PSFs between Figure 1D and 1E is intended as the focal length and cubic phase strength change with $d$. The PSFs, however, remain nearly wavelength-invariant for a fixed displacement, validating the behavior of the Alvarez EDOF metalens. To quantify the spectral bandwidth of our imaging system, we perform memory effect correlation calculations[47] as a function of input wavelength for our PSFs and estimate that our PSF similarity ranges from ~440-640 nm (see supporting information and Figure S8). Similarly, we assess the input angle dependence of our PSFs to estimate a field of view[48] of ~36° for our system (see supporting information and Figure S7). In our calculations, we assume an incoherent imaging model in which the system is linear in intensity and the intensity PSF is the modulus squared of the coherent PSF determined via angular spectrum propagation. While a coherent imaging system with such a device is theoretically possible, in practice it would be challenging due to speckle and a coherent source would require detection not only of intensity but also phase information for deconvolution, which is significantly more challenging at optical frequencies.

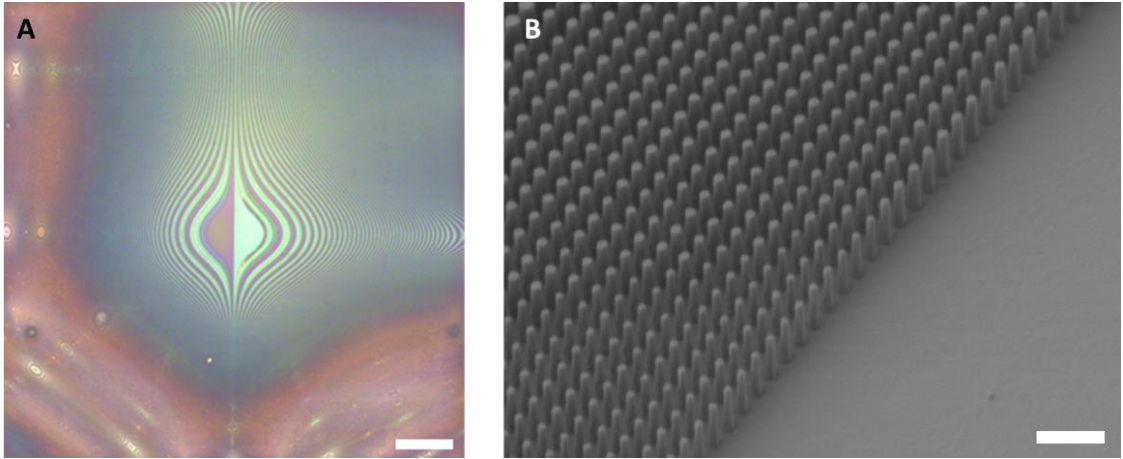

Figure 2: **Fabrication** (A) An optical image at normal incidence of one of the Alvarez EDOF metalens' plates is shown. Scale bar 0.125 mm. (B) A scanning electron micrograph captured at 45° depicting the fabricated silicon nitride nanoposts. Scale bar 1 µm.

*Experimental Verification:*

We fabricated the metasurfaces and an optical image and a scanning electron micrograph of one of the completed structures is shown in Figure 2A and 2B respectively. We experimentally verified the varifocal behavior of the metalens by measuring its response under 530 nm illumination (see Figure S5 for a schematic of the experimental setup). Figure 3 depicts the intensity along the $x = 0$ plane after the metalens for values of $d$ ranging from 25 µm to 125 µm. For each tuning state, the focal spot exhibits an extended depth, and the z-position of maximum intensity for each state agrees well with the theoretical focal length of a conventional Alvarez metalens, i.e., for $B = 0$ (Figure 3 inset). As both the quadratic and cubic terms in equation (4) are proportional to $d$, the depth of focus also increases with increasing focal length.

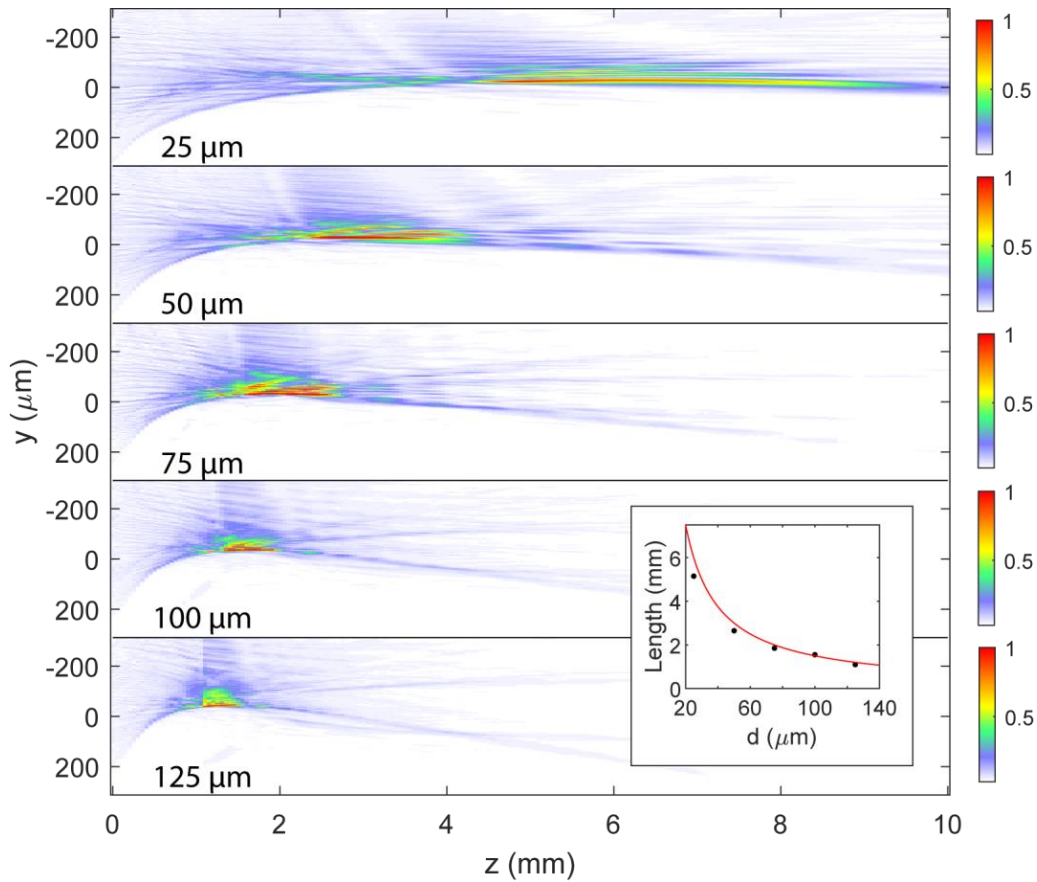

Figure 3: **Varifocal Behavior** Normalized measured intensity cross sections along the optical axis of the Alvarez EDOF metalens are depicted for different values of lateral displacement ranging from 25 µm to 125 µm going from the top to the bottom. The inset depicts the z distance to the point of maximum measured intensity (black dots) compared to the theoretical focal length of an Alvarez lens without an EDOF (red curve).

We next examined the chromatic focal shift of the device by illuminating with blue and red light and capturing intensity cross sections in the vicinity of the desired focal plane. Figure 4A (4B) shows the intensity for the case of a 100 μm (50 μm) lateral displacement about the 1.5 mm (3 mm) plane. While the longitudinal shift of the extended focal spot is significant over the wavelength range measured, for all three wavelengths the spatial intensity distributions at the desired focal plane (indicated by the dashed lines) are quite similar, elucidating the mechanism of generating spectrally invariant PSFs with an EDOF Alvarez metalens. The tested device also achieved an average diffraction efficiency of 37% at the three test wavelengths over the 25 µm to 125 µm actuation range (Figure S4).

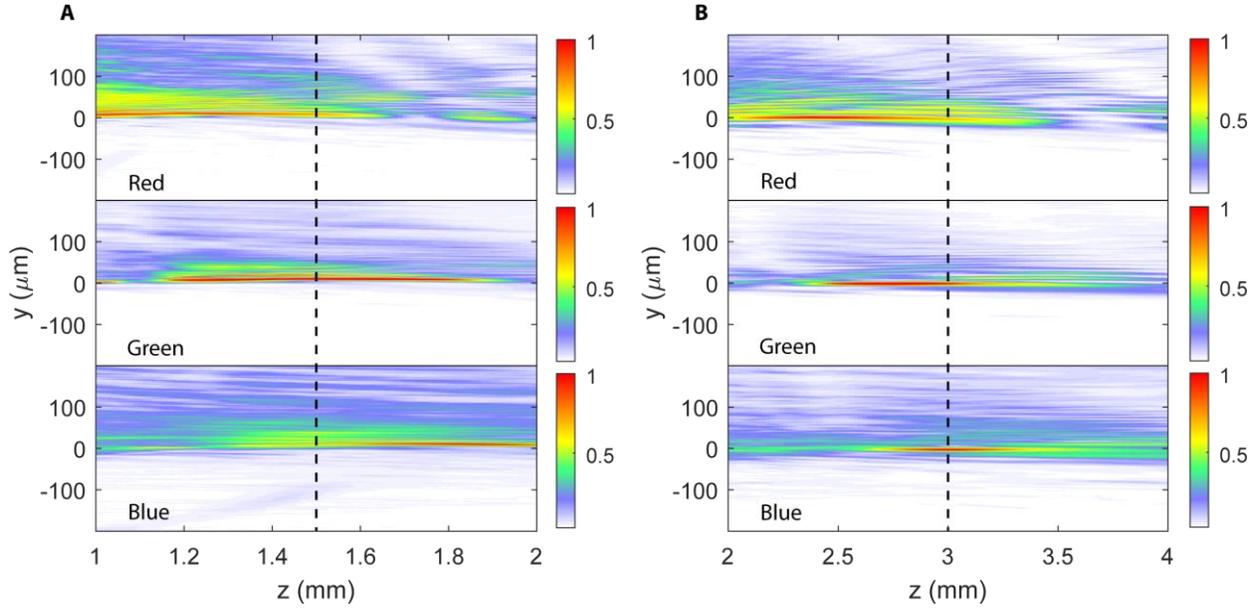

Figure 4: **Chromatic Focal Shift** Normalized measured intensity cross sections are shown for lateral displacements of 100 µm (A) and 50 µm (B) about their respective focal planes for red (625 nm), green (530 nm), and blue (455 nm) illumination cases.

To verify the spectral invariance of the system, we measured the PSFs (Figure 5) for different values of $d$ and compared to those of a static singlet metalens with a phase given by

$$\varphi_{singlet}(x, y) = \frac{2\pi}{\lambda}\left(f - \sqrt{x^2 + y^2 + f^2}\right), (8)$$

We fabricated two singlets with 1 mm apertures, 1.5 mm and 3 mm focal lengths, for green light (wavelength 530 nm) and their PSFs are shown in Figure 5A-C and 5H-J respectively. While these elements produce a tightly focused PSF for the green design wavelength, they vary drastically at other wavelengths, generating large diffraction blurs under red and blue illumination. For the case of the Alvarez EDOF metalens, however, its PSF remains much more similar under these different illumination wavelengths as shown in Figure 5D-F (5K-M) for a displacement of 100 µm (50 µm) measured at the same focal plane position as the 1.5 mm (3 mm) singlet. MTFs corresponding to these PSFs are shown in Figure S3, allowing us to estimate the experimental resolution of the system in terms of the 10% point for the MTF averaged over wavelength, giving 11 cycles/mm and 9.6 cycles/mm for the 3 mm and 1.5 mm focal length states respectively.

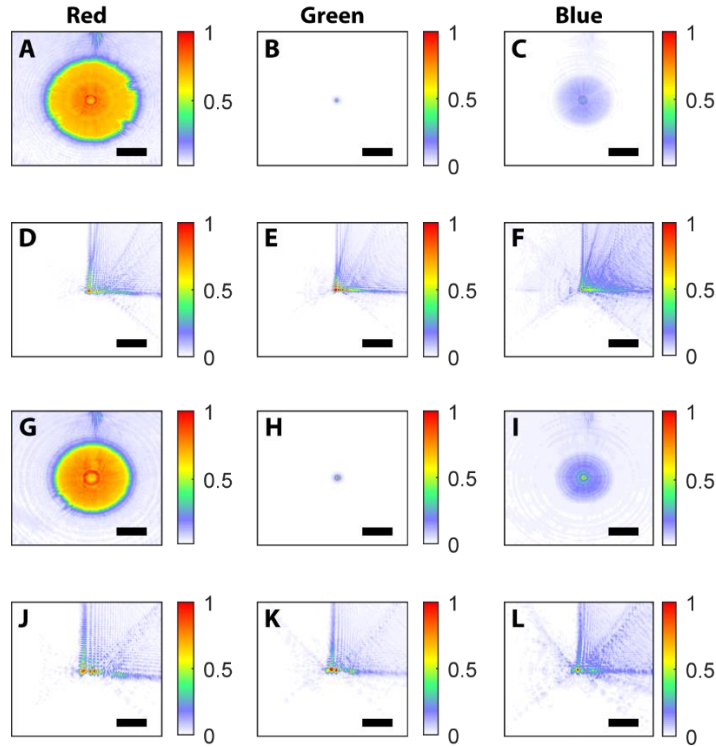

Figure 5: **Experimental Point Spread Functions** PSFs for the 1.5 mm singlet are shown (A-C) compared to those at an equivalent focal length for the Alvarez EDOF metalens with a lateral displacement of 100 µm (D-F). In (G-I) equivalent PSFs are depicted for the 3 mm singlet, while (J-L) show the corresponding Alvarez EDOF metalens' PSFs under 50 µm displacement at a 3 mm image plane. Scale bar 78.3 µm.

*Imaging:*

Leveraging the wavelength insensitivity of the Alvarez EDOF metalens' PSF, we performed imaging experiments (see Figure S6 for a schematic of the experimental setup) by capturing and deconvolving images of object patterns under white light illumination. Our image model can be compactly summarized via the equation $y = Kx + n$, where $y$ denotes the vectorized blurry captured image, $K$ is a matrix form of the measured PSF, $n$ is noise, and $x$ is the latent image that we want to reconstruct. Various deconvolution algorithms have been employed to solve this problem in computational imaging systems; here we elect to solve for $x$ using a regularized approach based on the total variation regularizer so that we can easily balance the tradeoff between deblurring and denoising, described via

$$x = argmin_x\ TV(x) + \frac{\mu}{2}\|Kx - y\|_2^2, (9)$$

In this equation, $TV$ denotes the total variation regularizer and $\mu$ is a hyperparameter that we can tune to adjust the weight assigned to deblurring or denoising[49]. In solving for $x$, we assume the PSF is spatially invariant, which reduces the complexity of equation (9), allowing us to use FFT operations rather than constructing the full $N \times N$ kernel matrix.

Applying this framework to our system, we imaged and deconvolved a set of three different object patterns with different underlying spatial features and color content (Figure 6) at five different magnification levels by tuning the lateral displacement of the metasurfaces. While a separate PSF is required for deconvolution in each tuning state, their near wavelength-invariant behavior enables us to use the same PSF for all colors. We also captured images of the same objects with our 1.5 mm singlet metalens for comparison. While much of the color content in the ground truth objects is smeared out by the large diffraction blurs of the singlet metalens, different colors are far more distinct for the deconvolved EDOF metalens images. Prior to deconvolution, the captured EDOF metalens' images are blurry for all colors, but the near spectrally invariant nature of the blur is the enabling condition for deconvolution with a single filter. Furthermore, while the singlet metalens is capable of imaging at only one magnification because of its static nature, in tuning the displacement of the Alvarez EDOF metalens from 25 $\mu m$ to 125 $\mu m$ and shifting the image plane, the magnification is seamlessly swept over a 5x zoom range.

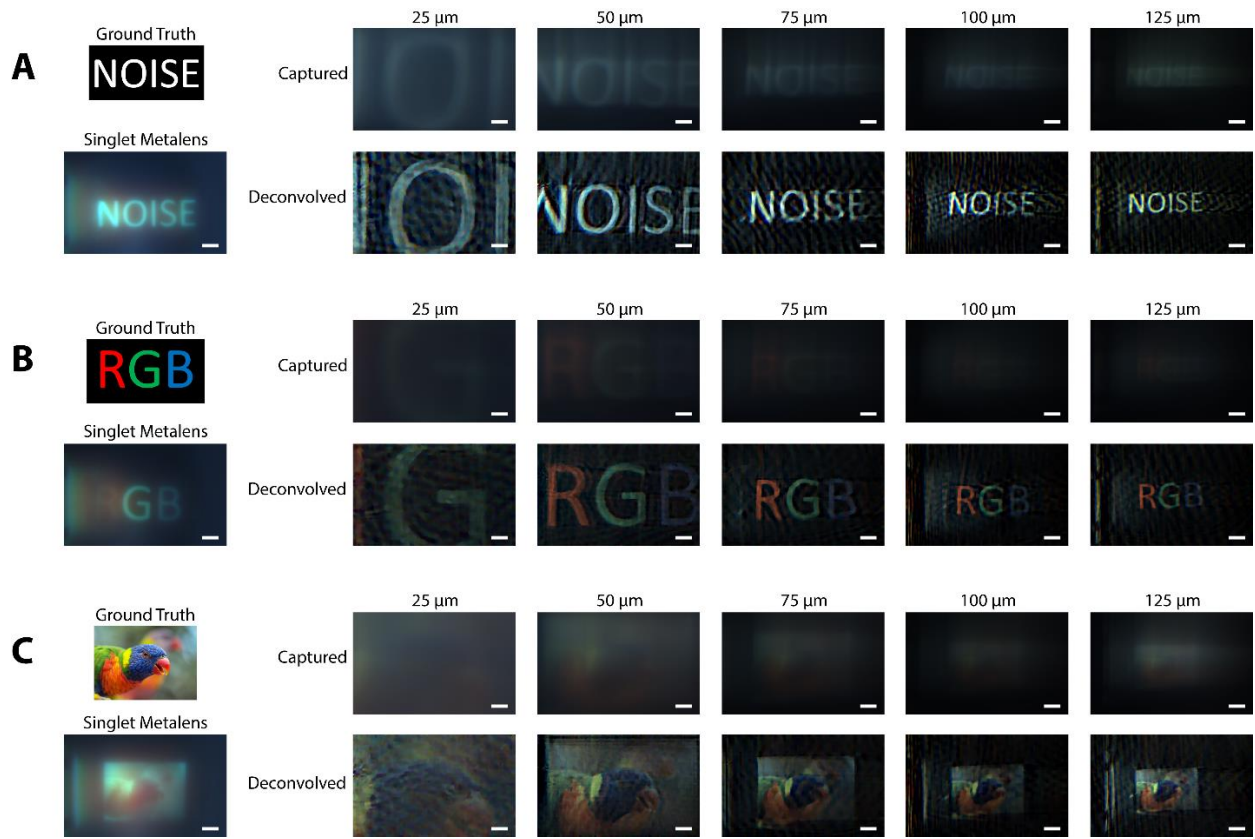

Figure 6: **White Light Imaging** Images are shown depicting the ground truth objects that were imaged, including "NOISE" text (A), "RGB" text (B), and a colorful bird (C). Also shown are images captured using the 1.5 mm singlet metalens under white light for comparison, and the images from the Alvarez EDOF metalens before and after deconvolution for five different lateral displacements ranging from 25 µm to 125 µm from left to right. Scale bar 62 µm in all images.

**Discussion:**

Our Alvarez EDOF metalens system demonstrates a wide tunable focal length range (a 4.8 mm or 400% change, $\frac{f_{max}-f_{min}}{f_{min}}$, or equivalently a 667-diopter change in optical power at 530 nm). This broad tuning range arises from the inverse proportionality of focal length to lateral displacement in equation (3). Simultaneously, the depth of focus of the lens is also tuned as the focal length shifts. This is not achieved by the inherent extension of the depth of focus that naturally comes with longer focal lengths, but by directly modifying the cubic phase strength via lateral displacement of a pair of phase elements, the first experimental demonstration of such behavior[40–42] to the best of our knowledge. For proof of concept, the metasurfaces are currently actuated by hand using kinematic stages, but for practical applications electrical control can be realized. Depending on the scale of the metasurface elements, a vast array of different actuation mechanisms may be employed. For devices with apertures on the order of hundreds of microns, for example, MEMS actuators could feasibly translate the metasurfaces, whereas for centimeter-scale devices, for which the masses and actuation distances are greater, a combination of miniature gears and stepper motors could be used[50].

While the tuning of the PSF is accomplished fully via mechanical actuation of the two plates, the zoom imaging behavior is unlike that of alternative tunable metalenses as a post-capture deconvolution step is required to produce a focused image. This requirement arises from the simultaneous wavefront coding that distinguishes our design from alternative varifocal metalenses. In applications with monochromatic or narrowband illumination, the added computational cost from the deconvolution step would introduce extra power consumption and latency that existing varifocal metalenses can avoid. As such, existing metasurface zoom lenses that do not depend on post-processing software for image formation would outperform our system in monochromatic applications. The added computation, however, extends the operating bandwidth and enables the system to image over a broad spectral range and not just with monochromatic light. As such, our system is well suited for imaging applications requiring a range of wavelengths and when additional power consumption and reduced SNR are tolerable.

Unlike metasurfaces optimized for operation at discrete wavelengths (e.g., at only red, green, and blue)[51], our system can produce in-focus images for intermediate colors such as yellow (Figure 6C). Compared to many of the existing broadband achromatic metasurface systems[19–22], our technique does not utilize specialized dispersion-engineered scatterers and as such does not have to contend with the area scaling limitations and polarization dependence. Furthermore, no other work exhibits a metasurface-based achromatic varifocal zoom. Our system, however, circumvents these challenges by sacrificing SNR to extend the depth of focus[30] and by adding post-processing software to the imaging pipeline. This adds power consumption to an imaging process that is conventionally passive and produces a latency that may hinder real-time operation when capturing video. Using a MATLAB implementation of the software on an ordinary personal laptop computer (Intel Core i7, 12 GB RAM), the deconvolution algorithm averaged 3.9 minutes per full-color 1936 × 1216 image. The deconvolution algorithm, however, could be hardware-accelerated by using field-programmable gate arrays (FPGAs) or adapted to run on state-of-the-art graphics processing units (GPUs) to enable real-time processing. Furthermore, in any application where real-time operation is unnecessary, all captured frames could be subsequently deconvolved offline. Alternative post-processing algorithms could also be used, including efficient and simple Wiener deconvolution, which averages 2.07 seconds per full-color image, representing a 113 × speedup, albeit at the cost of greater noise amplification and ringing artifacts in the absence of TV regularization (see Figure S9). Depending on the required magnification level, these artifacts can become more pronounced but depending on the requirements of the system, the speedup offered may outweigh the costs.

While the deconvolved images from our device exhibit noise amplification, these features can be mitigated with improvements in design and noise calibration. The limited aperture of our metalens (1 mm) precludes a high photon count and constrains our imaging to low light levels, contributing to the noise. With a larger aperture and increased space-bandwidth product, however, the Alvarez EDOF metalens would be capable of imaging with higher fidelity[3]. Additionally, in this work we have not performed any sophisticated noise calibration beyond dark noise subtraction. By calibrating and calculating the Poisson and Gaussian noise components of the sensor[52], the deconvolution and denoising could be significantly improved.

**Conclusion:**

In this article, we demonstrated the design, fabrication, and characterization of the first polarization-independent metasurface-based achromatic imaging system with varifocal zoom. We derived the phase functions for the metasurfaces in an Alvarez lens-like configuration that enable a varifocal lens with an extended depth of focus, and then experimentally demonstrated a 400% change in focal length with a nearly spectrally invariant PSF across the visible regime at each focal length. Leveraging our tunable focal length EDOF design, we imaged object patterns in full color over a 5x zoom range with significantly mitigated chromatic aberrations compared to conventional metalenses. While our captured images exhibit noise, we anticipate improvements in this via more advanced sensor calibration[52] and utilizing a wider aperture design[3]. With the demonstrated imaging and the ease in scaling our device to various sizes, our work may benefit a variety of applications including planar cameras, microscopy, augmented reality systems, and autonomous navigation.

**Acknowledgements:**

This research was supported by the Royalty Research Fund from the University of Washington (UW), a Samsung GRO grant, and the UW Reality Lab, Facebook, Google, and Huawei. Part of this work was conducted at the Washington Nanofabrication Facility / Molecular Analysis Facility, a National Nanotechnology Coordinated Infrastructure (NNCI) site at the University of Washington, which is supported in part by funds from the National Science Foundation (awards NNCI-1542101, 1337840 and 0335765), the National Institutes of Health, the Molecular Engineering & Sciences Institute, the Clean Energy Institute, the Washington Research Foundation, the M. J. Murdock Charitable Trust, Altatech, ClassOne Technology, GCE Market, Google and SPTS. We would also like to acknowledge useful

**Supporting Information Available:**

Materials and Methods

Transmission coefficient calculations at multiple wavelengths

Justification of the local phase approximation

Modulation transfer function measurement

Figures S1-S6

This material is available free of charge via the Internet at http://pubs.acs.org